\documentclass[a4paper]{article}

\usepackage{INTERSPEECH2019}
\usepackage{array}
\usepackage{color}
\usepackage{multirow}
\usepackage{url}

\title{LSTM Language Models for LVCSR in First-Pass Decoding and Lattice-Rescoring}
\name{Eugen Beck$^{1,2}$, Wei Zhou$^{1,2}$, Ralf Schl\"uter$^1$, Hermann Ney$^{1,2}$}
\address{
  $^1$Human Language Technology and Pattern Recognition, Computer Science Department,\\
  RWTH Aachen University, 52074 Aachen, Germany\\
  $^2$AppTek GmbH, 52062 Aachen, Germany}
\email{\{beck, zhou, schlueter, ney\}@cs.rwth-aachen.de}

\begin{document}

\maketitle
\begin{abstract}
LSTM based language models are an important part of modern LVCSR systems as they significantly improve performance over traditional backoff language models. Incorporating them efficiently into decoding has been notoriously difficult. In this paper we present an approach based on a combination of one-pass decoding and lattice rescoring. We perform decoding with the LSTM-LM in the first pass but recombine hypothesis that share the last two words, afterwards we rescore the resulting lattice. We run our systems on GPGPU equipped machines and are able to produce competitive results on the Hub5'00 and Librispeech evaluation corpora with a runtime better than real-time. In addition we shortly investigate the possibility to carry out the full sum over all state-sequences belonging to a given word-hypothesis during decoding without recombination.
\end{abstract}
\noindent\textbf{Index Terms}: speech recognition, decoding, LSTM language models, lattice rescoring

\section{Introduction}
In recent years, language models (LMs) based on long short-term memory (LSTM) neural networks have become an integral part of many state-of-the-art automatic speech recogition systems \cite{Xiong2018,Saon2017,Kyu2018}. LSTMs thus supersede traditional backoff-models which are based on word counts. For count based models relative frequencies of word n-grams are computed and stored. Due to the sparseness of the training data many n-grams are not seen and probability mass has to be allocated to them by smoothing out the probability distribution, e.g. by Kneser-Ney smoothing \cite{Kneser1995}. During decoding this sparseness can be utilized to recombine word-end hypotheses. LSTM LMs on the other hand use an internal state that is updated for each seen word to produce word posterior probabilities. This, in theory, gives them unlimited context that has to be considered which leads to large increases in required computation time.

In this work we show how to deal with this problem from a decoding point of view. We have chosen to utilize General Purpose Graphics Processing Units (GPGPUs) as an inference platform. This is reasonable, as more and more dedicated co-processors for machine learning workloads become available. On the server-side, GPGPUs by nVidia/AMD, TPUs by Google and (soon) AWS Inferentia-chips provide highly parallel computation platforms. Even for low-power devices chips like Edge TPU (by Google) or Kirin 970 (by Huawei) provide highly parallel computation platforms. We show that using an LSTM-LM in 1-st pass decoding is better than rescoring of lattices generated with a backoff LM. In addition forcing recombination of histories that share a trigram context during the 1st pass followed by lattice rescoring yields the same WER at lower RTF.

This paper is organized as follows: First we give an overview of existing work. Afterwards we present some implementation details. This is followed by a description of the models, corpora and hardware used for our experiments. Then we present a series of experiments on the Switchboard and Librispeech corpora.

\section{Related Work}
Using Neural Network based Language Models (NN-LMs) in Decoding is computationally more expensive than using backoff Language Models. In this section we give a short overview of how other researchers have dealt with this problem.

Early approaches of introducing NN-LMs into decoding include some form of conversion to a more traditional backoff LM: A very straightforward approach to convert complex models is to sample them to create large training corpora on which back-off LMs can be trained on. This is the approach of \cite{Deoras2011Approximation}. In \cite{Lecorve2012} the continues states of an RNN-LM are discretized to create a weighted finite state transducer. The authors of \cite{Arisoy2014} trained feed-forward LMs for different orders and extracted the probabilities for the backoff LM directly from the neural network. \cite{Adel2014} compares different techniques for conversion and \cite{Singh2017} uses these techniques to investigate conversion of domain adapted LSTM LMs.

Another option is to reduce the number of operations required for NN-LMs. For models with large vocabulary the lion's share of the computations occurs in the final layer, where a large matrix multiplication is required. For models using the softmax activation function the probability for all words needs to be computed even if only the probability for one word is required. Thus there is a large interest in developing techniques to avoid this. One popular approach is Noise-Contrastive-Estimation \cite{Gutmann2010,Gutmann2012}. Noise-Contrastive-Estimation is an adaptation of the loss function used in training to guide the model into a state where the output before the softmax is already approximately normalized and thus only a dot product is required to compute the probability of one word. It is used by \cite{Chen2015, Sethy2015, Huang2017}. In \cite{Huang2017} other methods like caching and the choice of activation functions within hidden layers are investigated aswell.

Yet another way to employ NN-LMs to improve ASR performance is to use them in a second pass for lattice rescoring. This is more efficient as only word sequences which are not pruned away need to be scored by the NN-LM. Examples for this approach are \cite{Deoras2011Rescoring, Liu2014, Sundermeyer2014, Liu2016, Kumar2017, Xu2018} where a variety of heuristics is proposed to speed up the rescoring process.

Closest to the work presented in this paper there are also publications where the authors integrated LSTM-LMs into first pass decoding: In \cite{Huang2014} a set of caches was introduced to minimize unnecessary computations when evaluating the LSTM-LM. In \cite{Hori2014}, an on-the-fly rescoring approach to integrate LSTM-LMs into 1-st pass decoding is presented. The authors of \cite{Lee2015} use a hybrid CPU/GPGPU architecture for real time decoding. The HCL transducer is composed with a small n-gram model and is expanded on the GPU while rescoring with an LSTM LM happens on CPU. Caching of previous outputs enables real-time decoding. All three papers use hierarchical softmax / word classes to reduce the number of computations in the output layer \cite{Morin2005} and with the exception of \cite{Huang2014} interpolate the LSTM-LM with a Max-Entropy LM \cite{Mikolov2011}. The works of \cite{Lee2015} are extended in \cite{Lee2018}. The LSTM Units are replaced with GRUs, NCE replaces the hierarchical softmax and GRU states are quantized to reduce the number of necessary computations.

\section{Implementation}
For this work we extended the decoder of the RWTH ASR toolkit, described extensively in \cite{nolden2017}. The decoder uses tree-conditioned search, which differs from the more common HCLG-based decoder in that we do not do static composition of the grammar WFST with the rest of the search network. Instead hypotheses from the HCL part of the decoder are grouped by their LM-history. Because these histories are opaque objects to the decoder we do not need to build a static WFST to represent the LM, which would be infeasible for LSTM-LMs anyway. Instead we only have to store the sequence of words and the state of the LSTM layers for each history. The language model itself can be any tensorflow graph as long as it is compatible with the general idea of a recurrent LM. The LM receives a state and one word and produces a new state and probabilities for the next word. One effect of statically combining Grammar and the HCL automaton is the pushing of the Grammar weights towards the start state of the transducer. In our decoder this early LM information is retrieved via Language-Model lookahead which is dynamically computed at runtime. We found that it is not necessary to use the LSTM-LM when computing this lookahead information to achieve the best possible WER. Only for very small beam-sizes we can observe a difference in WER. For our rescoring experiments we use push-forward rescoring as described in \cite{Sundermeyer2014}.

\section{Experiments}

\subsection{Hardware and Measurement Methodology}
Each node used for our experiments has two sockets with Intel Xeon E5-2620 v4 CPUs with a base-clock speed of 2.1Ghz and 4 Nvidia Geforce 1080Ti GPUs. Unless stated otherwise, our decoder ran in a single thread. The tensorflow runtime spawns more threads as it sees fit. As we are primarily using the GPU to do computations we set \textrm{intra/inter\_op\_parallelism\_threads} to 1. To compute the real time factor (RTF) we measure the total wallclock time required by the recognizer/rescorer to process all segments within the corpus and divide it by the total duration. This includes loading features from disk, forwarding them through the acoustic model and decoding / rescoring. Startup time is not included. Features are not extracted on the fly as it creates higher load on our fileserver and is not a major part during decoding anyway. In a research context preextracting features for a common task is useful as they are required for many experiments. In a production streaming system, feature extraction can be offloaded into a separate thread and will only contribute to latency, but not (significantly) to RTF.

\subsection{Corpora and Models}
In this paper we present results on two tasks. The first task is the 300h Switchboard-1 Release 2 which is evaluated on the Hub5’00 corpus. The second corpus is Librispeech \cite{panayotov2015}.

All our systems use 40 dimensional Gammatone features \cite{Schlueter2007}. The acoustic model is a multilayer BLSTM neural network trained with the state-level minimum Bayes Risk (sMBR) criterion \cite{Gibson2006}. The output units of the acoustic models are tied triphone states obtained using a Classification and Regression Tree (CART). When we use an LSTM-LM we interpolate it with the backoff-LM using log-linear combination. The perplexities of the models can be found in Table~\ref{tab:ppl}. The sizes of all models can be found in Table~\ref{tab:modelsize}. More information about the Librispeech system can be found in \cite{luescher2019}.

\begin{table}
\begin{center}
\begin{tabular}{|c|c|r|r|r|r|}
  \hline
  Corpus                                             & Model & Topology & \#out & \#param \\ \hline
  \multicolumn{1}{|c|}{\multirow{2}{*}{Switchboard}} & AM    &    6x500 &    9K &  47.6M  \\ \cline{2-5}
                                                     & LM    &   2x1024 &   30K &  82.8M  \\ \hline
  \multicolumn{1}{|c|}{\multirow{2}{*}{Librispeech}} & AM    &   6x1000 &   12K & 152.5M  \\ \cline{2-5}
                                                     & LM    &   2x2048 &  200K & 486.8M  \\ \hline
\end{tabular}
\end{center}
  \caption{Sizes of acoustic (AM) and langugae models (LM) used in this paper. The format for the topology column is \#layers times \#units}
\label{tab:modelsize}
\end{table}

\begin{table}
\begin{center}
\begin{tabular}{|l|r|r|}
  \hline
  Corpus                                                  & +LSTM-LM &    PPL \\ \hline
  \multicolumn{1}{|c|}{\multirow{2}{*}{Hub 5'00}}         & no       &  79.45 \\ \cline{2-3}
                                                          & yes      &  50.94 \\ \hline
  \multicolumn{1}{|c|}{\multirow{2}{*}{Librispeech dev}}  & no       & 146.18 \\ \cline{2-3}
                                                          & yes      &  70.59 \\ \hline
  \multicolumn{1}{|c|}{\multirow{2}{*}{Librispeech test}} & no       & 151.81 \\ \cline{2-3}
                                                          & yes      &  73.96 \\ \hline
\end{tabular}
\end{center}
\caption{Perplexities of Language Models used. Backoff models optionally combined with LSTM}
\label{tab:ppl}
\end{table}

\subsection{Baseline}
\begin{figure}
  \input{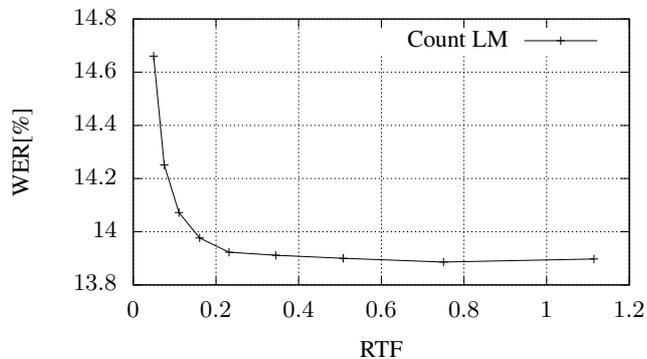}
  \caption{Backoff-LM baseline}
  \label{fig:count}
\end{figure}

Our baseline uses a 4-gram count model with an sMBR trained acoustic model. It saturates at a WER of 13.9\% with a RTF of 0.23. A detailed WER/RTF plot can be found in Figure~\ref{fig:count}.

\subsection{Parallelism}
\begin{figure}
  \input{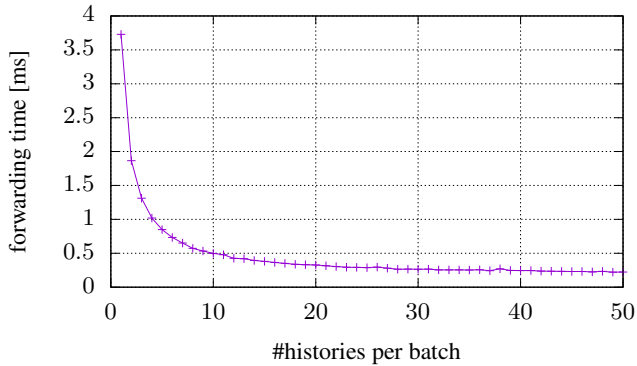}
  \caption{Time to process one batch for one step with the Switchboard LM divided by the number of histories in the batch}
  \label{fig:parallel}
\end{figure}

As GPGPUs are massively parallel architectures it is important to provide them with enough opportunities for parallelization when doing computations. In Figure~\ref{fig:parallel} we show the time it takes to forward one batch for various batch sizes divided by the number of histories for the Switchboard LM, i.e. we divide the total computation time by the number of histories within one batch. We can clearly see that it is much more efficient to forward many histories at the same time. Thus, once the LM receives the request to compute a specific word probability that is not already computed, we look for other histories that are not forwarded yet. We prioritize those histories that have a hypothesis close to a word-end state and with a score close to the currently best hypothesis.

\subsection{Effect of recombination}
The biggest difference between traditional backoff models and LSTM Language Models from the decoding point of view is the fact that backoff LMs allow for recombination of hypotheses at word-ends. For the LSTM-LM this is not possible, as each sequence of words for the history vector forms a unique history, that in principle can encode all words hithero seen. But of course the state of LSTM-Layers is finite and thus the LSTM will not be able to store an arbitrary amount. Furthermore some information about the past context might not be relevant for the search process anyway. Thus, it is reasonable to assume that even for an LSTM LM we can recombine hypotheses if the last $n$ words match. To empirically determine $n$ we conducted experiments in which we recombined word-end hypothesis where the last $n$ words matched. We did not recompute the state of the LSTM for this reduced context, but kept the context of the word end with the lowest score (i.e. highest probability). This forced recombination changes the lattice structure from a tree back to a directed graph.

\begin{figure}
  \input{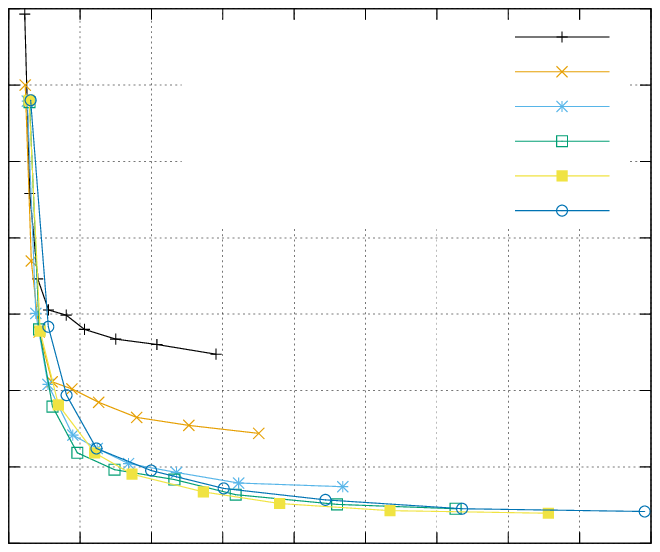}
  \caption{Different recombination limits for the Switchboard system}
  \label{fig:recombination}
  \vspace{-0.6cm}
\end{figure}

In Figure~\ref{fig:recombination} we show the results of our experiments. We can see that for a recombination limit of 5 we already reach the best WER (11.7\%) when rounding to 1 digit, as is usual for this task. Larger beam sizes do not yield significant improvements in terms of best achievable WER, but they allow us to reach it faster or get better WER for a fixed RTF value. This is because for $n=5$ we need a larger beam to reach the best possible WER, while larger $n$ allow for smaller beam sizes. For RTF $\sim$~1 a recombination of $n \ge 9$ should be selected.

\subsection{One-pass vs Two-pass with Rescoring}
\begin{figure}
  \input{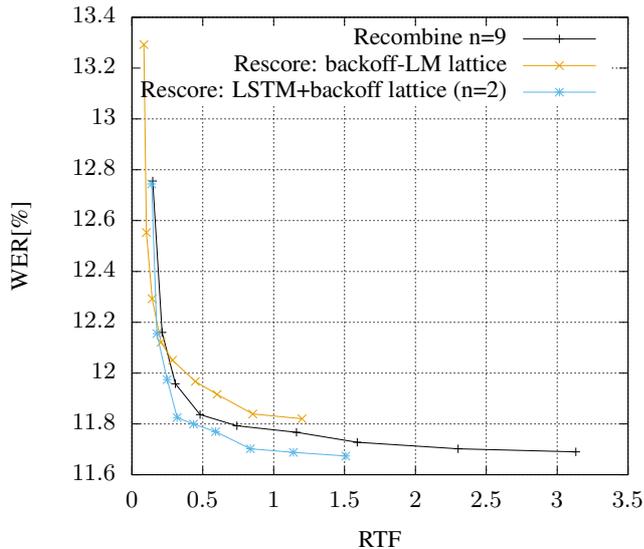}
  \caption{Comparison of first-pass recognition, lattice rescoring of backoff-models and lattice rescoring of LSTM-LM based lattices}
  \label{fig:rescoring}
  \vspace{-0.3cm}
\end{figure}

As a next step, we want to measure the performance of our system when the LSTM-LM is applied during lattice rescoring. As a baseline, we rescore the lattices generated by decoding with the backoff-LM. As can be seen in Figure~\ref{fig:rescoring} this is very fast for small beam sizes. For larger beam sizes it is still faster than using the LSTM-LM in the first pass, but it yields slightly worse results (11.8\% vs 11.7\%). In theory, if the beam-pruning were big enough, the lattice would contain the same word sequences that receive the best scores from the LSTM-LM. However this does not happen for the usual beam sizes we use during recognition as Figure~\ref{fig:rescoring} shows. As lattice rescoring is relatively fast (0.02-0.12 RTF, depending on lattice size) we also use it in combination with one-pass decoding with LSTM-LMs. In order to keep the complexity of the first pass low we use a small recombination threshold ($n = 2$). This will produce lattices with many recombinations and thus many opportunities for lattice rescoring to find new word-sequences. Our experiments show that this indeed is more efficient than performing one-pass decoding with a high recombination limit.

\subsection{Full-sum decoding}
The one pass decoding without recombination makes it possible to apply full sum over all state sequences in the HMM instead of Viterbi approximation in the Bayes decision rule. We also explore this effect on the Hub 5'00 data set. We use the same decoder framework as implementation baseline and changed the auxiliary function of dynamic programming to be the probability summation over all incoming paths instead of maximization. The pruning threshold of the beam search has to be set a bit higher to keep most of the contributing paths. Paths of pronunciation variants are normalized and merged as well. The simple hand-crafted time distortion penalties are normalized back to probability domain. Grid search is used to find the optimal acoustic and language model scaling. To our surprise, although mathematically more accurate, the full-sum decoding leads to the same accuracy w.r.t single best in the lattice, while the size of search space is generally larger. However, applying confusion network decoding on the lattice additionally further improves the full-sum result from 11.7\% to 11.4\%, while no difference is obtained for Viterbi decoding. We will further investigate the effect of full-sum decoding in future work.

\subsection{Librispeech}
We conducted initial experiments on Librispeech but did not yet complete a full analysis. Our results can be found in Table~\ref{tab:librispeech}. Due to the significantly larger vocabulary compared with the Switchboard system (200k vs 30k) our absolute RTF are much higher when using the LSTM-LM in one-pass mode (if one wants to get the best possible WER). The best performing strategy on Switchboard of using an LSTM LM in first pass with a short recombination limit and rescoring the resulting lattice is again the best performing strategy at a RTF of one. However the best WER is only reached at an RTF far above one. This could be alleviated by training LSTM LMs using Noise Contrastive Estimation \cite{Gutmann2012}. This is reserved for future work. Another observation of interest here is that decoding on the \textit{other} condition takes much longer than on the \textit{clean} condition. This is not too surprising as decoding an utterance where the acoustic model is less confident will yield more scores that are close together. Here the difference in WER is a factor of around 2, while the difference in RTF is a factor of around 4.

\setlength{\tabcolsep}{0.4em}
\begin{table}
  \begin{tabular}{|l|r|r|r|c|r|r|}
  \hline
    Corpus                                            & 1-st pass LM                                       & Recmb. &                               Rescr. LM &  WER &   RTF \\ \hline
    \multicolumn{1}{|c|}{\multirow{4}{*}{dev-clean}}  & \multicolumn{1}{c|}{\multirow{2}{*}{backoff}}      &    N/A &                                       - & 3.72 &  0.56 \\ \cline{3-6}
                                                      &                                                    &    N/A &                            interpolated & 2.55 &  0.97 \\ \cline{2-6}
                                                      & \multicolumn{1}{c|}{\multirow{2}{*}{interpolated}} &     10 &                                       - & 2.39 &  6.95 \\ \cline{3-6}
                                                      &                                                    &      2 &                            interpolated & 2.40 &  4.06 \\ \hline
    \multicolumn{1}{|c|}{\multirow{2}{*}{dev-other}}  &                                            backoff &    N/A & \multicolumn{1}{c|}{\multirow{5}{*}{-}} & 8.73 &  2.25 \\ \cline{2-3} \cline{5-6}
                                                      &                                       interpolated &   none &                                         & 5.75 & 13.38 \\ \cline{1-3} \cline{5-6}
    \multicolumn{1}{|c|}{\multirow{2}{*}{test-clean}} &                                            backoff &    N/A &                                         & 4.18 &  0.56 \\ \cline{2-3} \cline{5-6}
                                                      &                                       interpolated &   none &                                         & 2.78 &  7.14 \\ \cline{1-3} \cline{5-6}
    \multicolumn{1}{|c|}{\multirow{2}{*}{test-other}} &                                            backoff &    N/A &                                         & 9.31 &  2.59 \\ \cline{2-3} \cline{5-6}
                                                      &                                       interpolated &   none &                                         & 6.23 & 15.62 \\ \hline
  \end{tabular}
  \caption{Results of various decoding strategies with backoff and LSTM-LM for the Librispeech dataset}
  \label{tab:librispeech}
  \vspace{-0.8cm}
\end{table}

\begin{figure}
  \input{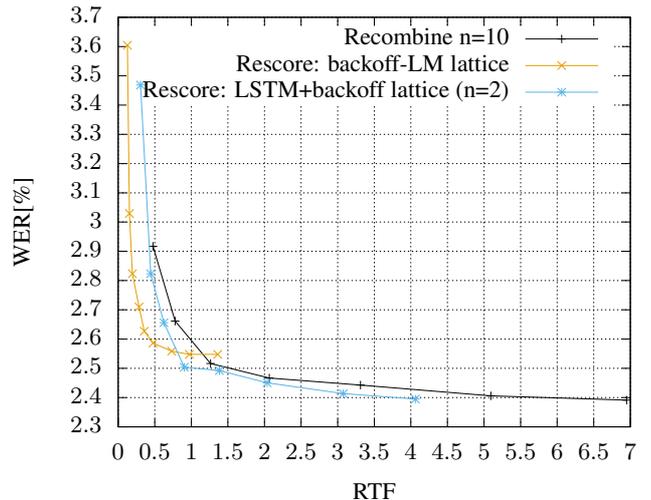}
  \caption{Comparison of first-pass recognition, lattice rescoring of backoff-models and lattice rescoring of LSTM-LM based lattices for the Librispeech dev-clean corpus}
  \label{fig:librispeech}
  \vspace{-0.6cm}
\end{figure}

\section{Conclusions}
In this paper we have shown how to use LSTM-LMs in decoding using a GPGPU. We have shown that first using the LSTM LM with a small recombination limit and doing lattice rescoring afterwards yields the most efficient decoding process. This approach yields a WER of 11.7\% on the Hub5'00 task at an RTF of 1. Further work is required for systems with very large vocabulary where the best possible WER is only reached for a RTF well above 1. Further improvements to the WER (11.7\% to 11.4\%) were obtained by full-sum decoding with subsequent confusion-network decoding.

\vspace{-0.2cm}

\section{Acknowledgments}
This project has received funding from the European Research Council (ERC) under the European Union’s Horizon 2020 research and innovation program (grant agreement No 694537, project "SEQCLAS") and from the European Union’s Horizon 2020 research and innovation program under the Marie Skłodowska-Curie grant agreement No 644283. The work reflects only the authors' views and the European Research Council Executive Agency (ERCEA) is not responsible for any use that may be made of the information it contains. Eugen Beck was partially funded by the 2016 Google PhD Fellowship for North America, Europe and the Middle East. We also want to thank our colleagues Kazuki Irie, Christoph L\"uscher and Wilfried Michel for providing us with the acoustic/language models.

\vspace{-0.3cm}

\bibliographystyle{IEEEtran}

\let\normalsize\footnotesize\normalsize

\let\OLDthebibliography\thebibliography
\renewcommand\thebibliography[1]{
  \OLDthebibliography{#1}
  \setlength{\parskip}{1.2pt}
  \setlength{\itemsep}{1pt plus 0.07ex}
}

\bibliography{mybib}

\end{document}